\pdfoutput=1 
\documentclass{JINST}
\usepackage{amsmath}                            

\usepackage{multicol}
\usepackage{verbatim}

\newcommand{\GERDA}{\textsc{Gerda}}
\newcommand{\GELATIO}{\textsc{Gelatio}}

\title{GELATIO: a general framework for modular digital analysis of high-purity
Ge detector signals}

\author{
M.~Agostini,$^{a}$
L.~Pandola,$^b$
P.~Zavarise$^{b,c}$ and
O.~Volynets$^d$\\
\llap{$^a$}~Physik-Department E15, Technische Universit\"at M\"unchen, \\
James-Franck-Str. 1, D-85748 Garching, Germany.\\
\llap{$^b$}~INFN Laboratori Nazionali del Gran Sasso, \\
SS 17-bis km 18+910, I-67100, Assergi (AQ), Italy.\\
\llap{$^c$}~Dipartimento di Fisica, Universit\`a dell'Aquila,\\
Via Vetoio Localit\`a Coppito, I-67100, L'Aquila, Italy.\\
\llap{$^d$}~Max-Planck-Institut f\"ur Physik, \\
F\"ohringer Ring 6, D-80805, M\"unchen, Germany.\\ \\
E-mail: \email{matteo.agostini@ph.tum.de}, 
\email{luciano.pandola@lngs.infn.it}, 
\email{paolo.zavarise@lngs.infn.it}, 
\email{volynets@mppmu.mpg.de}\\
}
\keywords{Software architectures (event data models, frameworks and databases),
Data processing methods, Gamma detectors (scintillators, CZT, HPG, HgI etc)}

\abstract{
\GELATIO\ is a new software framework for advanced data analysis and digital signal
processing developed for the \GERDA\
neutrinoless double beta decay experiment.
The framework is tailored to handle the
full analysis flow of signals recorded by high purity Ge detectors and
photo-multipliers from the veto counters. It
is designed to support a multi-channel modular and flexible analysis, widely
customizable by the user either via human-readable initialization files or via a
graphical interface.  
The framework organizes the data into a multi-level structure, from the raw data
up to the condensed analysis parameters, and includes tools and
utilities to handle the data stream between the different levels. 
\GELATIO\ is implemented in C++. It relies upon \textsc{Root} and its extension
\textsc{Tam}, which provides compatibility with PROOF, enabling the software to
run in parallel on clusters of computers or many-core machines.
It was tested on different platforms and benchmarked in several 
\GERDA-related applications. A stable version is presently available for
the \GERDA\ Collaboration and it is used to provide the reference analysis of
the experiment data.} 

\begin{document}
\section{Introduction}
The ``GERmanium Detector Array'' (\GERDA) is an experiment looking 
for neutrinoless double beta decay of $^{76}$Ge which is presently under 
commissioning at the underground Laboratori Nazionali del Gran Sasso of INFN, 
Italy~\cite{gerda,gerda2}. The neutrinoless double beta decay is a process which
violates by two units the lepton number conservation and is forbidden in the
Standard Model. 
Its experimental observation would imply the Majorana nature of the neutrino,
also providing a first measurement of the neutrino absolute mass scale. 
The experiment will use an array of high-purity germanium (HPGe) detectors 
isotopically enriched in $^{76}$Ge (about 40~kg are planned for the final 
configuration), aiming to achieve a substantial reduction in 
the background at the $Q_{\beta\beta}$-value of the $^{76}$Ge decay with 
respect to the predecessor experiments~\cite{hm,hm2,igex}. 

The background reduction in \GERDA\ is obtained by an innovative design approach
in which naked HPGe detectors are operated directly in ultra radio-pure liquid
argon which acts as coolant material and as  passive shielding against the external
radiation. The cryogenic liquid is surrounded by an additional thick layer of 
ultra-pure water, which is effective in shielding external 
neutrons and $\gamma$-rays. The water volume is instrumented with photo-multipliers 
and is operated as a Cherenkov detector to reject events due to high-energy muons. 
Part of the remaining background events can be identified by analyzing the HPGe signal shapes
and applying pulse shape discrimination (PSD) techniques~\cite{dusan, agomc, psdigex, psdhm}.
Moreover, the \GERDA\ collaboration is testing in the R\&D set-up called
\textsc{LArGe}~\cite{large} the liquid argon instrumentation with
photo-multipliers (PMTs) which would provide an active veto system.

In the present commissioning phase, an array of three non-enriched HPGe detectors 
has been deployed in the \GERDA\ set-up and is in data taking.
The charge signals from the HPGe detectors are sampled by 14-bit Flash-ADCs
(FADC) running at $100\,\text{MHz}$ sampling rate and stored to disk for off-line analysis.
For each physical trigger all the HPGe detector signals are
acquired to check for coincidences. A second data stream of the experiment is
provided by the PMT signals from the Cherenkov muon veto, 
which are digitized by the same FADCs used for the HPGe detectors.

In this paper a data analysis framework called \GELATIO\ (GErda LAyouT 
for Input/Output) is presented and discussed. The framework was 
developed to handle the full data analysis flow 
of \GERDA\ as well as of all the R\&D activities related to the experiment.
The framework has been designed to be solid, user-friendly, flexible, 
maintainable over a long lifetime and scalable to the future phases of the 
experiment. Furthermore, thanks to its generic interfaces, it could 
be used in other activities involving off-line 
analysis of digitized pulses from HPGe detectors or other kinds of
detectors. 

The paper is organized as follows.
In section~\ref{sec2} the main requirements driving the software design
and the basic concepts of the framework are presented in detail. 
Section~\ref{sec3} describes the software implementation and the technical
solutions pursued.
A few examples about the validation and the application of the framework 
are reported in section~\ref{sec4}.
A summary and discussion of future plans are eventually presented in the final
section.

\section{Concept and design} \label{sec2} 
\GELATIO\ is a data analysis framework designed to provide a flexible
environment and a complete suite of tools for off-line digital signal
processing and for analysis of data recorded with HPGe detectors. The
software aims to provide a common platform for the \GERDA\ Collaboration to run
the analysis of the experiment data.
To meet these requirements the framework must be able to:
\begin{itemize}
\item decouple the algorithm implementation from the raw data format 
allowing the users
   to run the same analysis algorithms on data sets acquired with different
   hardware and\,/\,or encoded in different formats; 
\item perform a modular and highly customizable digital signal processing. This
   approach aims to simplify and foster as much as possible the re-use, the
   sharing and the comparison of the analysis algorithms, avoiding unnecessary
   duplications in the code and improving its validation;
\item optimize the computational performances and be cross-platform
   compatible, hiding the technical aspects to the end users.
\end{itemize} 
The solution worked out is based on two paradigms which are discussed below: 
multiple level data organization and modular digital
signal processing.

\subsection{Multi-level data structure} \label{req1}
The raw data, the information extracted by the signal processing
and the analysis results are stored in a hierarchical structure. 
This approach aims to increase flexibility and enables a multi-user customized 
data analysis. Alternative analyses can be created as forks of the 
default one, sharing part of the data flow until a given level and then 
creating a parallel stream of information. 
The multi-level structure includes naturally in the framework
the conversion of the raw data into a new standardized format which is optimized 
for signal
processing and data storage. After the conversion, all data
can be processed along the same analysis stream independently 
of the parent data acquisition (DAQ) system 
data format, including data produced by Monte Carlo simulations.

\begin{figure}[tp]
   \begin{center}
      \includegraphics[width=\textwidth]{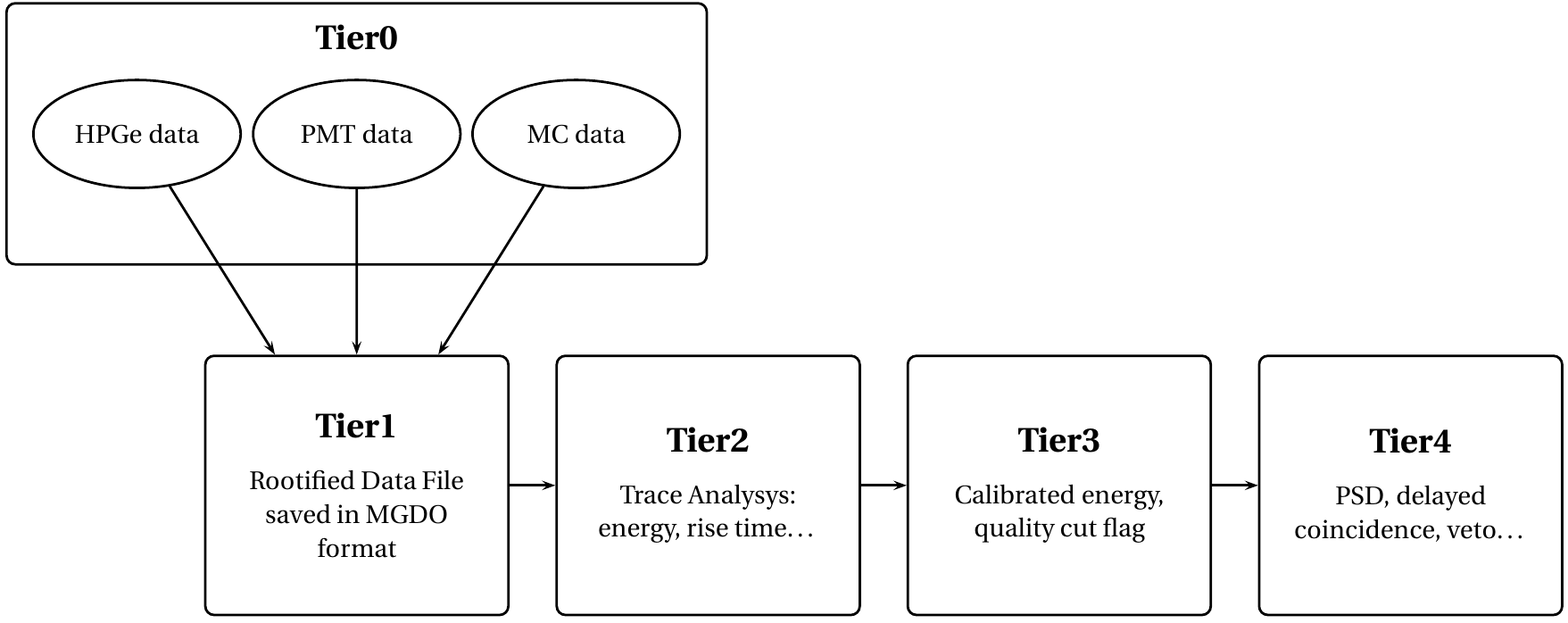}
   \end{center}
   \caption{The hierarchical organization of the data in \GELATIO.
   The framework organizes the output of each step of the analysis in a
   different level (Tier) starting from the raw data (Tier0) up to the condensed
   parameter of the final analysis. The Tier1 contains the same information of
   the raw data but encoded with a different format based on
   \textsc{Root}~\cite{root} and \textsc{MGDO}. 
   More details can be found in sect.~3.2.
   } 
   \label{tiers} 
\end{figure}

The multi-tier structure is depicted in \figurename~\ref{tiers}.
The raw data provided by the different DAQ systems and by the Monte Carlo 
simulations are stored in the lowest level (``Tier0''). 
Data are then converted into a new encoding and stored as ``Tier1''. The first
two tiers contain exactly the same amount of information, the only difference
being that while the Tier0 is the native DAQ format, the Tier1 is a standardized 
format that can be chosen to be solid, flexible, exportable and easily readable. 
The Tier1 data 
are distributed to the \GERDA\ collaborators as the starting point for the
analysis. 
Higher-level tiers -- which are produced from the Tier1 -- 
are meant to contain the 
analysis results. The ``Tier2'' files store the output information 
obtained by applying the digital analysis to the individual traces of 
each event, as rise time, amplitude, average noise, baseline average value, etc.
Similarly, the ``Tier3'' files store information extracted from the Tier2, e.g. the 
actual energy spectrum obtained by calibrating the amplitude spectrum with the 
appropriate calibration curves.
As the analysis becomes more and more refined (noise rejection, pulse shape
discrimination analysis, delayed coincidence, veto, etc.), the information can be 
stored in higher-level tiers. 


A drawback of this approach is the additional request for disk space due to the
coexistence of the same information in both the Tier0 and the Tier1.
On the other hand the raw data are not meant to be distributed because the
Collaboration plans to blind the events with energy close to
the region of interest ($Q_{\beta\beta}$-value of $^{76}$Ge).
Raw data will be backed up in the computing centers of three different
institutions and used only to generate the Tier1.
In the conversion process the data blinding is applied and only 
the resulting Tier1 data are released to the Collaboration.
\subsection{Modular digital signal processing}\label{dsp}
The core of \GELATIO\ is the digital signal processing which creates the Tier2
files starting from the detector signals stored in the Tier1. 
In this step different algorithms are applied to the signals in order to extract
efficiently the pulse shape information, for instance maximum amplitude,
rise time, baseline slope, etc.
In $\gamma$-ray spectroscopy these operations are usually performed by chains of
elementary digital filters (differentiation, integration, deconvolution, etc.)
optimized to reduce the noise and to extract the information with high precision.

To support a highly customizable analysis, the design of \GELATIO\ is based on a modular approach. 
The analysis is divided into modules, each handling a unique and consistent task
of the digital data processing, as for instance energy reconstruction and
baseline subtraction.  
Each module includes a chain of elementary digital filters which is optimized to
extract the information of interest from the signal trace. The computed information
as well as the shaped traces can be used as input for other modules.
The list of active modules and the parameters used by the internal chain of filters are
controlled by the end user through an appropriate ASCII initialization file (INI file). 

\begin{figure}[t]
   \begin{center}
      \includegraphics[width=\textwidth]{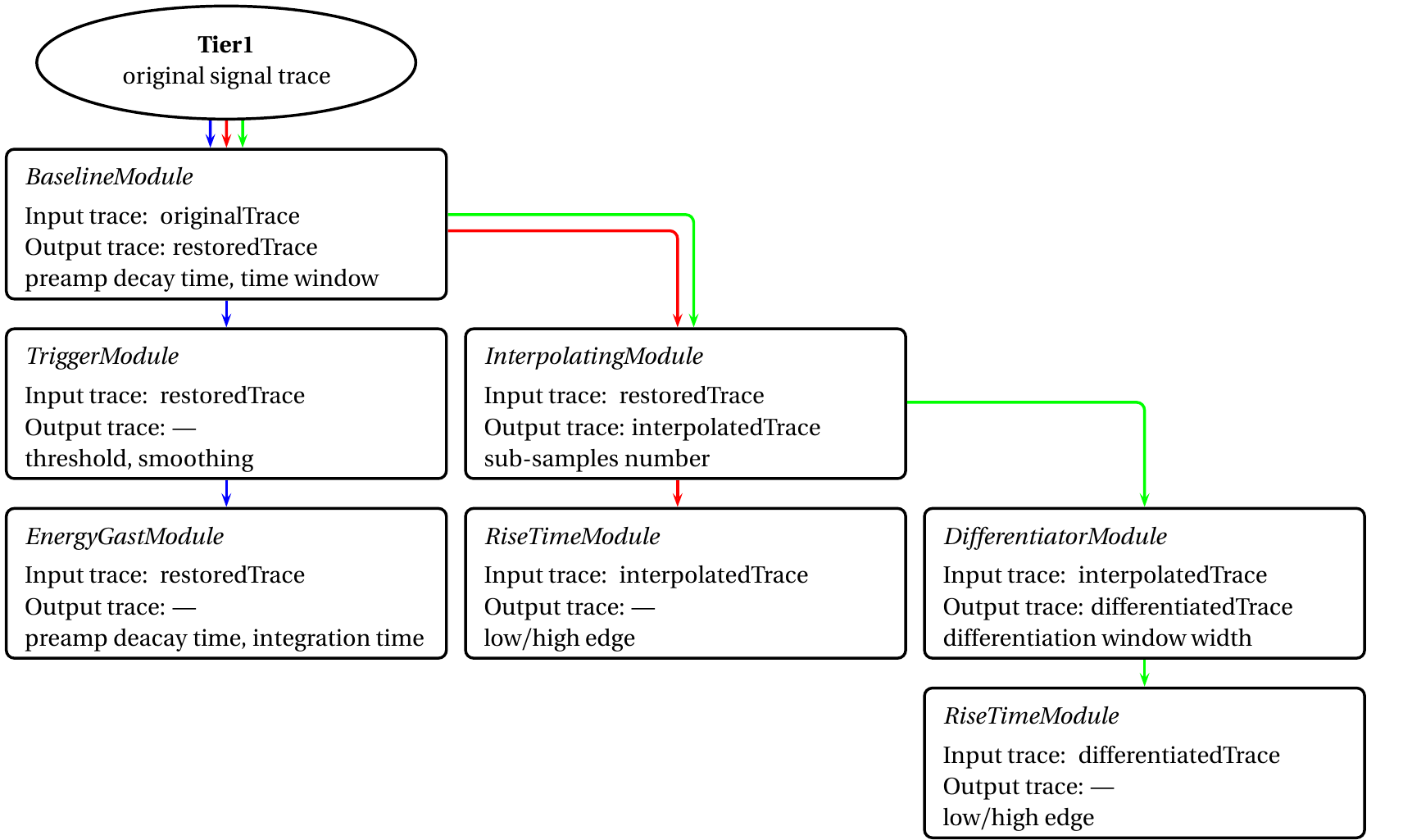}
   \end{center}
   \caption{
   Data flow of an illustrative analysis which uses three chains of modules.   
   The first chain (blue arrows) reconstructs the event amplitude and
   includes the baseline restoration and correction for pile-up
   (\emph{BaselineModule}) and the computation of the trigger time
   (\emph{TriggerModule}).
   The signal shaped by \emph{BaselineModule} and the trigger computed by
   \emph{TriggerModule} are used as input for \emph{EnergyGastModule}, which reconstructs
   the pulse amplitude according to the Gast moving-window-deconvolution
   approach~\cite{gast}.
   The second chain (red arrows) is used to estimate the rise time of the
   signal. The traces are first shaped by \emph{BaselineModule}, interpolated by
   \emph{InterpolatingModule} to push the time resolution below the
   sampling frequency and finally processed by \emph{RiseTimeModule}. 
   The last chain (green arrows) computes the rise time of the derivative of the
   signals (current signal) and shares the first three modules with the previous
   chain. The signal shaped by \emph{InterpolatingModule} is fed to
   \emph{DifferentiatorModule} to compute the numerical derivative, and eventually is
   parsed to a second instance of \emph{RiseTimeModule}.
   The figure shows for each module the input and output trace (second and third
   line) and the main parameters used by the internal algorithms
   (last line).
   }
   \label{dataflow}
\end{figure}

This design provides a wide flexibility as complicated chains of modules can be
created by the user in an open and transparent way through the INI file. 
The same module can be run many times within the same execution and used in
different chains, each time with different sets of parameters.
Moreover, the user can easily create new modules implementing his own customized
analysis tasks. The new modules are immediately available for registration in 
the INI file and can be combined with the standard ones to create new chains.
The data flow and the INI file of an illustrative analysis
are reported in \figurename~\ref{dataflow} and \figurename~\ref{inifile}, 
respectively.
This solution enhances also the re-use of the analysis algorithms and avoids
code proliferation. 
\begin{figure}[t]
   \begin{center}
      \begin{multicols}{2}
      \footnotesize
\begin{verbatim}
[Parameters]
FileList=tier1_r01.root tier1_r02.root
OutputFile=output.root
LogFile=output.log

[TaskList]
Task_BaselineModule_1=true
Task_TriggerModule_1=true
Task_EnergyGastModule_1=true
Task_InterpolatingModule_1=true
Task_RiseTimeModule_1=true
Task_DifferentiatorModule_1=true
Task_RiseTimeModule_2=true

[Task_BaselineModule_1]
InputTraceName=originalTrace
OutputTraceName=restoredTrace
BaselineRestorationStart=100ns
BaselineRestorationStop=2us
TauPreamp=47us
PileUpCorrection=true

[Task_TriggerModule_1]
InputTraceName=restoredTrace
NumberOfSigmaThs=3
TimeAboveThs=100ns
IntegrationWindowWidth=50ns
[Task_EnergyGastModule_1]
InputTraceName=restoredTrace
DifferentiationWindowWidth=10us
IntegrationWindowWidth=8us
FlatTopPosition=0.4

[Task_InterpolatingModule_1]
InputTraceName=restoredTrace
OutputTraceName=interpolatedTrace
SubSampleNumber=10

[Task_RiseTimeModule_1]
InputTraceName=interpolatedTrace
LowEdge=10
HighEdge=90

[Task_DifferentiatorModule_1]
InputTraceName=InterpolatedTrace
OutputTraceName=differentiatedTrace
DifferentiatonWindowWidth=50ns

[Task_RiseTimeModule_2]
InputTraceName=differentiatedTrace
LowEdge=10
HighEdge=90

\end{verbatim}
      \end{multicols}
   \end{center}
   \caption{Example of an INI file implementing the analysis described in
   \figurename~2. The INI file is organized in blocks. The first two blocks
   (Parameters and TaskList) are used to define the input and output files and
   to register the list of modules, respectively. The following blocks are used
   to define the parameters of the registered modules, as for instance the input
   and output traces.}
   \label{inifile}
\end{figure}
%

\section{Implementation} \label{sec3} 
The core of the framework is implemented in C++ to ensure an easy and natural
interfacing with several scientific general-purpose projects.
This choice provides also high computational performances, wide flexibility thanks to
the object-oriented programming support, and cross-platform compatibility.
\GELATIO\ depends on the CLHEP~\cite{clhep} and FFTW3~\cite{fftw3} libraries 
for scientific computing, and on the \textsc{Root}~\cite{root} and 
\textsc{Tam}~\cite{tam} 
libraries for the management of the modular analysis, the data storage and 
the graphical tools. All the external software packages 
mentioned above are freeware and open-source. \GELATIO\ 
additionally depends on the \textsc{MGDO} package for the basic digital 
signal processing algorithms and for the definition of the objects used to 
encapsulate the information in the Tier1 output. \textsc{MGDO} 
(Majorana-\GERDA\ Data Objects) is a set of libraries that are jointly 
maintained and developed by the Majorana~\cite{majorana} and \GERDA\ collaborations. They 
are specifically designed to improve the encapsulation and the handling of complex 
data as dedicated C++ objects. 

\GELATIO\ is distributed to the \GERDA\ collaborators in the form of a source
code. It can be compiled on any platform supporting GNU C++,
including Linux and MacOS. A \texttt{configure} script 
takes care of setting automatically the appropriate paths and environment 
variables necessary
to compile the code. 
The installation procedure has been successfully tested 
on both 32- and 64-bit operating systems. 

To ensure flexibility and good computational performances, the framework 
includes both compiled and interpreted code. 
In section~\ref{t0t1} and section~\ref{t1t2} the implementation of the two
executables in charge for the raw data to Tier1 conversion and for the actual
modular data analysis (Tier2 production) are described in detail.
Then the suite of Bash and Python scripts to handle the data streaming through
the different tiers is presented in section~\ref{MU}.
Section~\ref{gui} eventually describes the graphical interface used to 
display the event traces, define the shaping parameters and create the INI files.
\subsection{Conversion of raw data to the analysis format}\label{t0t1}
The binary data format chosen for Tier1 is a \textsc{Root} file 
containing a \texttt{TTree} of \textsc{MGDO} objects (\texttt{MGTEvent} and \texttt{MGTRun}). 
The \textsc{MGDO} objects employed in the Tier1 output are containers which 
encapsulate the basic information of individual events (signal traces, time stamps, 
DAQ flags, etc.) and of runs (start and stop times, run type). 
The usage of a \textsc{Root} files has many advantages, most notably the streamers of the
\textsc{Root} objects, the compression routines and the interface to the
\textsc{Root} graphic
utilities.

The conversion of raw data in the Tier1 format is performed by the executable 
\texttt{Raw2MGDO}, which accepts a list of raw data files and lets the user 
customize the name and the number of the output files.
The framework contains dedicated classes (``Decoders'') which are used by
\texttt{Raw2MGDO} to decode the supported binary raw files, in order to read the
information to be copied to the Tier1 structure. At the moment, six different 
decoders are available in \GELATIO, supporting all data formats currently  
employed in the \GERDA\ activities.  
The decoders take care of extracting the information from the raw file and of all 
the required preprocessing -- as endianness inversion -- before writing them in 
the \textsc{Root} file. The \GELATIO\ decoders inherit by the same virtual base class, in 
order to improve flexibility and to avoid code duplication. The common interface 
defined by the virtual base class eases the extension\,/\,upgrade of the present 
decoders as well as the implementation of new decoders to read any other 
binary data format. 

Such an approach makes the framework able to handle in a completely 
transparent way a data stream containing an unspecified number of DAQ channels, 
each with digitized traces of unspecified length. This is required because the 
number of operational detectors and the digitization parameters will 
change during the experiment lifetime. Similarly, \GELATIO\ is able to handle a mixed 
stream coming from different types of detectors (e.g. HPGe detectors and PMTs in 
\textsc{LArGe}). The Tier1 data format is used also as output of the pulse simulation 
software developed in the framework of \GERDA~\cite{agomc}. 
Consequently, the simulated traces can be treated in the same way of the
experimental data and be processed along the same analysis flow,
enabling an easy and direct Monte Carlo-to-data comparison.

The testing and benchmarking of \GELATIO\ was performed by using the main
$\GERDA$ server. The server runs Scientific Linux 5.5 64-bit and mounts
a Dual Xeon E5620 CPU
(2$\times$4 cores at 2.4~GHz with 2$\times$12~Mb cache), 16~GB of RAM,
and 20 hard-disks (2~TB) connected through a SATA~3\,Gb/s interface and
operated in RAID6/XFS.
The computational performances of the conversion program are affected by the
encoding and type of raw data and by the \textsc{Root} compression options required.
For instance, a typical \GERDA\ calibration run 
(about $3.5\cdot 10^6$ waveforms, each having 4096 samples, total size about 30~GB) 
is completely converted into the Tier1 format by the reference machine in about
100~minutes using a single thread. 
The processing time is substantially reduced if the \textsc{Root} compression option is
switched off, at the expense of additional disk space. It has to be 
emphasized that the conversion of raw data into the Tier1 format must be 
done only once, so the best compromise is usually to pay in CPU computing 
time to obtain a smaller output. 



\subsection{Implementation of digital signal processing}\label{t1t2}
To implement the modular analysis following the design 
discussed in section~\ref{dsp}, the framework relies on the Tree-Analysis 
Module (\textsc{Tam}) package.
\textsc{Tam}~\cite{tam} is a free package for \textsc{Root} developed 
to provide a very general and modular
interface for analyzing data stored in a \texttt{TTree}.
The software combines the features of two \textsc{Root} objects: the \texttt{TSelector} method for
processing trees and the \texttt{TTask} for handling a hierarchical structure of modules 
in a user-transparent way. 

In \GELATIO\ each analysis module is a concrete class derived by the basic interface
\texttt{TAModule} provided by \textsc{Tam} via an additional \GELATIO-specific 
base class named \texttt{GERDAModule}. 
\textsc{Tam} is used to handle the event
loading from the Tier1 file, the exchange of information among different modules
and the object output list. 
Moreover, the interfacing with \textsc{Tam} ensures the compatibility with the 
\textsc{Root} extension PROOF (Parallel ROOT Facility)~\cite{proof} enabling the software to  
run several threads in parallel.

The executable in charge of the Tier2 creation takes care to instantiate the
\textsc{Tam} interface -- initializing the analysis modules according to the instruction
provided through the INI file -- and to store all the outputs of the same execution
in a single \textsc{Root} file.
The output is a collection of \textsc{Root} objects usually containing a \texttt{TTree}
for each module but also histograms or signal traces.
The software provides also a master \texttt{TTree} which can be used
for unrestricted and parallel access to information contained in any other
\texttt{TTree} in the file, via the \textsc{Root} friendship mechanism.

The CPU time required to run an analysis depends on the number of active channels
and modules as well as on the module parameters.
For instance, the standard \GERDA\ analysis chain includes baseline restoration, 
trigger position and rise time computation, and two independent modules for 
amplitude reconstruction.
A typical calibration run containing $3.5 \cdot 10^6$ traces, 
each 4096 samples long, is processed according to the \GERDA\ standard analysis chain 
in less than 4 hours by using a single thread of the reference benchmark
machine.
%
\subsection{Utilities}\label{MU}
To help the handling of the data stream through the different tiers, \GELATIO\
includes a suite of utilities implemented as Bash and Python scripts.
The utilities work over a well-defined directory structure (``analysis file
system'') and provide a user-interface for each step of the analysis. 
The scripts take care of identifying which files should be processed and of storing the
results of each step in the proper directory, including a log file collecting
the standard outputs generated by the executables.
Since the information is stored in fixed directories inside the file system,
each user is immediately able to recover any output.

\begin{figure}[t]
   \begin{center}
      \includegraphics[width=\textwidth]{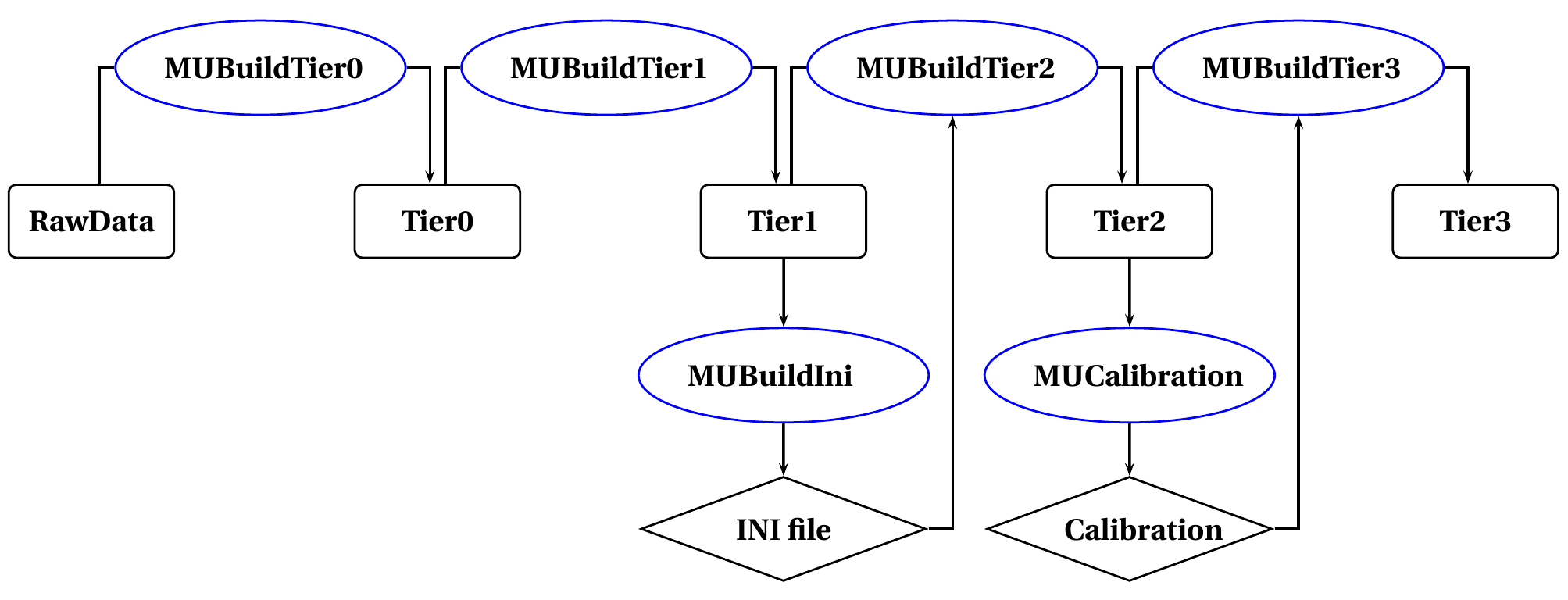}
   \end{center}
   \caption{Data flow of the information through the different tiers performed by using
   the suite of utilities (blue nodes) included in \GELATIO. 
   These utilities are designed to provide an interface to import raw data in
   the framework, convert them into the Tier1 format, create the INI file and run
   the digital signal processing.
   Moreover, the utilities allow the user to perform an interactive and graphical
   calibration of the amplitude spectra.}
   \label{muflow}
\end{figure}

In \figurename~\ref{muflow}, each step of the data flow is depicted together with the
associated utility.
The utilities up to the Tier2 builder are implemented as simple Bash scripts and
are designed to provide an interface to the file system and to run the \GELATIO\
executables with the proper options. 
The last two routines are more complicated as they are supposed to provide
to the user an interactive graphical tool to calibrate the energy spectra and to create the
Tier3. Moreover, they take care of storing the calibrating parameters of each channel
in different directories of the output \texttt{TFile}, together with all the
information important for the debugging, i.e. the calibration log files and the
plots of the fits.
These utilities are implemented in Python to take advantage of the \textsc{Root} binding
\texttt{PyROOT} which enables cross-calls from Python to ROOT/CINT~\cite{root}.

\subsection{The graphical interface} \label{gui}
\begin{figure}[t]
  \[
  \begin{smallmatrix}
    \includegraphics[trim=0pt 8pt 0pt 0pt, clip]{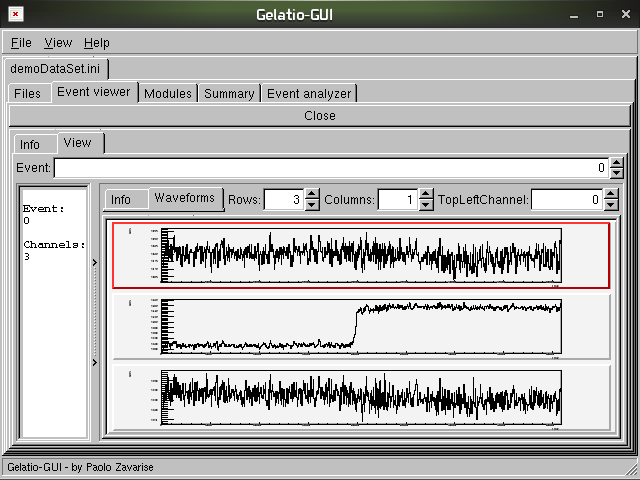}&
    \includegraphics[trim=0pt 8pt 0pt 0pt, clip]{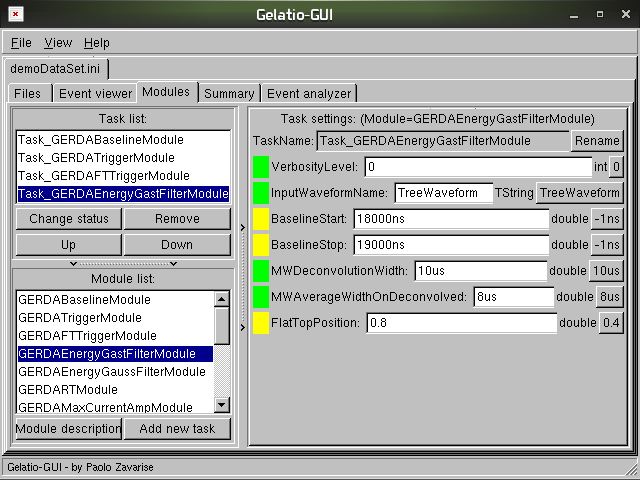}\\
    ~\\
    \text{(a)} & \text{(b)}\\
    ~\\
    \includegraphics[trim=0pt 8pt 0pt 0pt, clip]{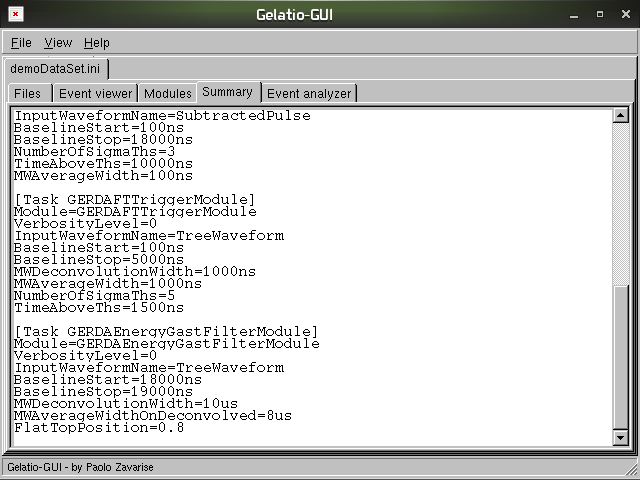}&
    \includegraphics[trim=0pt 8pt 0pt 0pt, clip]{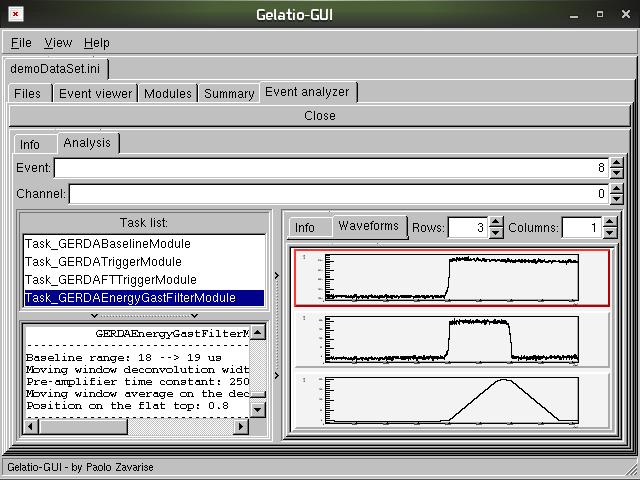}\\
    ~\\
    \text{(c)} & \text{(d)}\\
  \end{smallmatrix}
  \]
  \caption{Screen shots of the \GELATIO\ GUI. The input Tier1 file comes
  from a \GERDA\ background run and contains three traces per event. The
  screen shots show the tools and utilities available in the GUI: (a) Event
  displayer. The signals from the three channels are displayed together. (b) INI
  file editor. It can be used to select and customize the analysis tasks to be
  performed. The ``Module list'' contains all the analysis modules available in
  \GELATIO.  (c) INI output summary. It shows the human-readable INI file
  produced according to the user choices in the INI file editor. (d) Event
  analyzer. To apply the full analysis chain to a given trace. The screen shot
  shows the intermediate shaped traces calculated  by the analysis module which
  implements the Gast~\cite{gast} algorithm for amplitude reconstruction.}
  \label{gelatiogui}
\end{figure}

The graphical user interface (GUI) integrated in \GELATIO\ (\figurename~\ref{gelatiogui})
is a general and powerful tool developed to create and handle INI files.
The interface is implemented entirely by using \textsc{Root} graphical
components.
Despite the lack of flexibility and the intrinsic limitations of the
\textsc{Root} graphical libraries, the native \textsc{Root} solution has been
preferred to ensure smooth integration with the rest of the framework and
to minimize external dependencies.  

The GUI aims to help the user in the creation and in the testing of the INI
files and is able to handle multiple files at the same time. The 
layout is based on five main tabs:
\begin{itemize}
\item{Files}: to select the Tier1 files to analyze. The files can be selected by using 
a graphical window, and multiple files can be selected at the same time.
\item{Event viewer} (\figurename~\ref{gelatiogui}.a): to browse channel-by-channel the events 
contained in the selected Tier1 files. Several channels can be displayed at the same 
time.
\item{Modules} (\figurename~\ref{gelatiogui}.b): to configure the tasks to be used 
for the analysis. A ``task'' is a module with a particular set of parameters. 
The modules to be activated can be selected from a list. For each module the GUI displays 
the full list of the customizable parameters which can be edited interactively by
the user. 
Since modules can be possibly configured with many independent parameters, it is particularly 
useful to have a visual list of them and of their default values. For each parameter 
the GUI shows a color-code label. The green code means that the selected value is equal to the 
default; the yellow code stands for a valid value which is different from the default; 
the red code indicates an invalid parameter value (wrong unit, for instance).
\item{Summary} (\figurename~\ref{gelatiogui}.c): to view the resulting INI file.
\item{Event Analyzer} (\figurename~\ref{gelatiogui}.d): to test the INI file, namely 
the analysis chain, on a single event in the Tier1 file. It is possible to display the 
input\,/\,output traces of each module and also the intermediate traces produced 
along the digital processing. Such a tool proved to be very useful for debugging the analysis 
chain and for tuning\,/\,optimizing the values of the parameters to be used for a given data set.
\end{itemize}

\section{Application and Benchmarking} \label{sec4} 
The framework has been used up to now to handle and analyze data 
from several \GERDA-related activities, including calibrations with 
radioactive sources. 
In particular, \GELATIO\ was used for the data analysis of the \GERDA\ R\&D
activities related to the Broad Energy Germanium (BEGe) detectors.
The experimental data~\cite{ago11,bu10,div09} and the corresponding 
Monte Carlo simulations~\cite{agomc} were processed on the same 
footing to compare directly the results. The pulse 
shape discrimination algorithms developed for the BEGe 
detectors~\cite{dusan,dusan2} were coded as dedicated \GELATIO\ modules 
and successfully applied to data. Results in term of  
discrimination efficiency calculated with the \GELATIO\ modules are 
consistent with those obtained with the dedicated code of 
Ref.~\cite{dusan}. 
The present analysis of the \textsc{LArGe} data~\cite{markThesis} is based on
\GELATIO; in this case, the framework is able to handle the data streams 
coming from the HPGe detector and from the PMTs of the 
instrumented liquid argon veto. Finally, 
the framework is currently used for the reference data analysis of the data 
collected in the \GERDA\ commissioning with three HPGe detectors. 

Up to now \GELATIO\ was used on data files coming from six  
independent DAQ systems, differing for binary data format, 
number of channels, sampling frequency and sampling window. 
It proved to be able to handle correctly  
multiple DAQ channels -- possibly referred to different kinds of 
detectors -- and pile-up corrections that must be applied for 
source calibration runs, because of the higher counting rate. The 
software turned out to be stable and robust. Fixes for the few 
minor bugs reported since the release of the stable version are made 
available in regularly-updated versions of \GELATIO.

\begin{figure}[tp]
   \begin{center}
      \includegraphics[width=\textwidth]{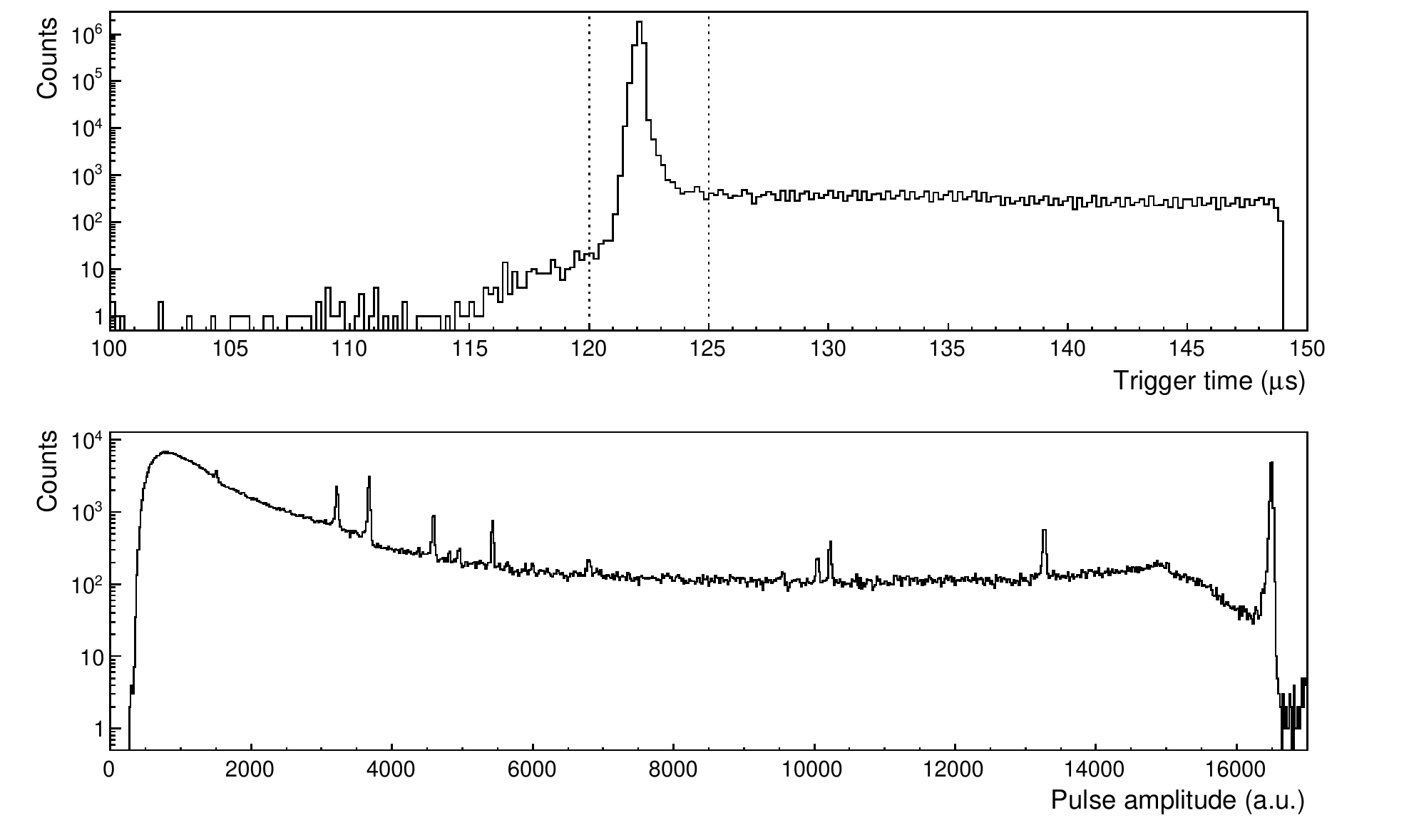}
   \end{center}
   \caption{Analysis results obtained by the \GELATIO\ 
processing of a $^{228}$Th calibration taken in \GERDA. Upper panel: 
distribution of the reconstructed trigger time for one of the three 
detectors. The DAQ chain is set in order to have the trace at 
about 120~$\mu$s after the start of the sampling window. Lower panel: 
amplitude distribution obtained for the same data set from the 
Gaussian shaping algorithm. The amplitude spectrum includes only the signals
having the trigger time between 120~$\mu$s and 125~$\mu$s (see dashed lines in
the upper panel) in order to discard accidental and mis-reconstructed events.
The management of the 
\GELATIO\ output is performed via the master \texttt{TTree} in the Tier2 
file.} \label{spettro}
\end{figure}
\figurename~\ref{spettro} shows the distribution of the trace 
trigger time (upper panel) and the amplitude spectrum (lower panel) 
of one of the HPGe detectors deployed in \GERDA\ irradiated with a 
$^{228}$Th calibration source. The distributions are obtained at the 
end of the \GELATIO-based analysis flow, using the reference INI file 
described in section~\ref{t1t2}. The amplitude spectrum of \figurename~\ref{spettro} 
has been produced after having discarded those events having a 
reconstructed trigger time significantly far from the expected value. Notice that 
trigger times and trace amplitudes are calculated by two 
independent analysis modules and stored in two separate \texttt{TTrees} in 
the Tier2 file. The amplitude and trigger information from the 
different \texttt{TTrees} is correlated via the master \texttt{TTree} created by 
\GELATIO.

The amplitude was reconstructed 
off-line using two independent algorithms -- Gaussian shaping (shown 
in \figurename~\ref{spettro}) and Gast method~\cite{gast} -- both implemented 
as \GELATIO\ modules. After the appropriate tuning of the parameters, the 
two methods give equivalent results for the energy resolution.
The energy reconstruction and resolution provided 
by \GELATIO\ for the \GERDA\ data were compared to the results
obtained with an independent and dedicated analysis code and found to be  
consistent. 

\section{Conclusions} \label{sec5} 
A powerful and flexible software framework called \GELATIO\ has been 
developed to handle the full analysis flow of the \GERDA\ experiment 
and the related R\&D activities. The software is written in 
C++ and is based on an object oriented design. \GELATIO\ 
contains executable programs and utility scripts which take care of 
the full analysis chain, starting from the raw data up to the final 
condensed parameters. Complex analysis, possibly 
involving energy reconstruction, detector coincidence and pulse 
shape discrimination, can be hence managed in a very transparent and 
general way. 

As raw data are converted in a standardized \textsc{Root}-based format,
data streams originated by the different DAQ systems used in the \GERDA\ and by
Monte Carlo simulations  can be treated with the same algorithms and along the
same analysis flow,
easing the cross-check and inter-comparison among the results.

A modular analysis approach based on \textsc{Tam} is used, designed to 
sub-divide the analysis work in many nearly-independent tasks that can 
be activated interactively and possibly run in parallel in multi-thread
systems. The active modules and the corresponding parameters are 
selected by the end user via a human-readable INI file or a graphical 
interface. The GUI acts also by event displayer and by interactive analysis
manager, able to show each intermediate step of the modular analysis. 

A stable version of \GELATIO\ is presently available for the \GERDA\ Collaboration
and the framework has been widely used for the data analysis in the \GERDA\ 
commissioning.
The \GERDA\ database system, which is currently under development, 
supports \GELATIO\ in input and output. It is able to
import the analysis results from Tier2 files and to generate Tier1 files from 
a custom selection of events made through \textsc{SQL} \cite{SQLref} queries.
\GELATIO\ is also used in other \GERDA-related activities, including \textsc{LArGe}
and the characterization of prototype BEGe detectors.

The software has been 
validated against other independent and dedicated analysis codes. Furthermore, 
\GELATIO\ proved to be robust, effective and flexible enough to handle a 
complex analysis stream from a real-life multi-channel experiment. \GELATIO\ 
could be used in any experimental activities involving digital 
pulse shape analysis of HPGe detector signals.

%
\acknowledgments
We would like to acknowledge our colleagues of the \GERDA\ Collaboration,
especially B.~Schwingenheuer, for many invaluable advices concerning 
analysis algorithms and data analysis, as well as for providing results from 
his own analysis code that were used to benchmark \GELATIO.
We want also to thank D. Bazzacco and C. A. Ur for all the stimulating
discussions concerning $\gamma$-ray digital filters and the BEGe team for having been our
beta tester and having provided very valuable feedback, in particular
D.~Budj\'{a}\v{s} and A. Lubashevskiy. 

We express our gratitude to the colleagues from the Majorana Collaboration, and
specifically J.~Detwiler and M. Marino, for pointing out to us the \textsc{Tam}
package and for many suggestions about software design and implementation.
Furthermore, we would like to thank S.~Stalio from the LNGS IT Service for help
and support with the LNGS computing cluster.

This work was supported in part by the Transregio Sonderforschungsbereich
SFB/TR27 ``Neutrinos and Beyond'' by the Deutsche Forschungsgemeinschaft and by
the Munich Cluster of Excellence ``Origin and Structure of the Universe''.



\begin{thebibliography}{99}

\bibitem{gerda} GERDA Collaboration, I.~Abt et al., \emph{GERDA:
The GERmanium Detector Array for the search of neutrinoless
$\beta \beta$ decay of $^{76}$Ge at LNGS}, Proposal,
\href{http://www.mpi-hd.mpg.de/ge76}{http://www.mpi-hd.mpg.de/ge76}.

\bibitem{gerda2} GERDA Collaboration, S.~Sch\"onert et al., 
\emph{The GERmanium Detector Array (GERDA) for the search of neutrinoless $\beta\beta$ 
decays of $^{76}$Ge at LNGS}, Nucl. Phys. B, Proc. Suppl. 
\textbf{145} (2005) 242.

\bibitem{hm} M.~Gunther et al., \emph{Heidelberg - Moscow beta beta experiment with Ge-76: Full 
setup with five detectors}, Phys. Rev. D \textbf{55} (1997) 54.
 
\bibitem{hm2} H.V.~Klapdor-Kleingrothaus, I.V.~Krivosheina, A.~Dietz, O.~Chkvorets, 
\emph{Search for neutrinoless double beta decay with enriched $^{76}$Ge in Gran Sasso 
1990-2003}, Phys. Lett. B \textbf{586} (2004) 198.

\bibitem{igex} C.E.~Aalseth et al., \emph{IGEX $^{76}$Ge neutrinoless double-beta 
decay experiment: Prospects for next generation experiments}, Phys. Rev. D
\textbf{65} (2002) 092007.

\bibitem{dusan} D.~Budj\'a\v{s}, M.~Barnab\'e Heider, O.~Chkvorets, N.~Khanbekov and 
S.~Sch\"onert, \emph{Pulse shape discrimination studies with a Broad-Energy Germanium 
detector 
for signal identification and background suppression in the GERDA double beta decay experiment}, 
JINST \textbf{4} (2009) P10007.

\bibitem{agomc}
M.~Agostini et al., \emph{Signal modeling of HPGe detectors with a small
read-out electrode and application to neutrinoless double beta decay search in 
Ge-76}, JINST \textbf{6} (2011) P03005.

\bibitem{psdigex}
D.~Gonzalez et al., \emph{Pulse shape discrimination in the IGEX experiment},
Nucl.\ Instrum.\ Meth.\ A {\bf 515} (2003) 634 [hep-ex/0302018].

\bibitem{psdhm}
J.~Hellmig, H.~V.~Klapdor-Kleingrothaus,
\emph{Identification of single-site events in germanium detectors by digital
pulse shape analysis}, Nucl.\ Instrum.\ Meth.\  A {\bf 455} (2000) 638.

\bibitem{large} 
M.~Di Marco, P.~Peiffer, S.~.Schonert,
\emph{LArGe: Background suppression using liquid argon (LAr) scintillation for 0
nu beta beta decay search with enriched germanium (Ge) detectors},
Nucl.\ Phys.\ Proc.\ Suppl.\  {\bf 172} (2007) 45 [physics/0701001].

\bibitem{root} R.~Brun and F.~Rademakers, 
\emph{ROOT - An Object Oriented Data Analysis Framework}, 
Nucl.\ Instrum.\ Meth. A \textbf{389} (1997) 81; URL 
\href{http://root.cern.ch/}{http://root.cern.ch/}.
%
\bibitem{gast} J.~Stein, F.~Scheuer, W.~Gast, A.~Georgiev, 
 \emph{X-ray detectors with digitalizer preamplifiers},
Nucl.\ Instrum.\ Meth. B \textbf{113} (1996) 141.

\bibitem{clhep} L.~L\"onnblad, \emph{CLHEP: a project for designing a C++
class library for high energy physics}, Comput. Phys. Commun. \textbf{84} (1994) 307;
URL \href{http://proj-clhep.web.cern.ch}{http://proj-clhep.web.cern.ch}.

\bibitem{fftw3} M.~Frigo and S.G.~Johnson, \emph{The Design and Implementation of FFTW3}, 
\emph{Proceedings of the IEEE} \textbf {93} (2005), 216; 
URL \href{http://www.fftw.org/}{http://www.fftw.org/}.

\bibitem{tam}
M.~Ballintijn, C.~Loizides and C.~Reed, \emph{Tree Analysis Modules}, 
URL \href{http://www.cmsaf.mit.edu/twiki/bin/view/Software/TAM}
{http://www.cmsaf.mit.edu/twiki/bin/view/Software/TAM}.

\bibitem{majorana} 
C.~E.~Aalseth {\it et al.} [ Majorana Collaboration ],
\emph{The Majorana neutrinoless double beta decay experiment},
Phys.\ Atom.\ Nucl.\  {\bf 67} (2004) 2002 [hep-ex/0405008].

\bibitem{proof}
M.~Ballintijn, R.~Brun, F.~Rademakers, G.~Roland, 
\emph{The PROOF Distributed Parallel Analysis Framework based on ROOT},
[arxiv:physics/0306110v1]; URL 
\href{http://root.cern.ch/drupal/content/proof}{http://root.cern.ch/drupal/content/proof}.



\bibitem{ago11} M.~Agostini et al., \emph{Characterization of a broad energy germanium 
detector and application to neutrinoless double beta decay search in 
$^{76}$Ge}, JINST \textbf{6} (2011) P04006.

\bibitem{bu10} M.~Agostini et al., \emph{Procurement, production and testing of BEGe detectors 
in $^{76}$Ge}, to appear in Nucl. Phys. B, Proc. Suppl. (Neutrino 2010).

\bibitem{div09} A.~di Vacri et al., \emph{Characterization of Broad Energy Germanium 
Detector (BEGe) as a candidate for the GERDA experiment}, 
IEEE Nucl. Sci. Symp., Conf. Record 2009 (2009) 1761.

\bibitem{dusan2} D.~Budj\'a\v{s}, \emph{Germanium detector 
studies in the framework of the GERDA experiment}, 
Ph.~D. thesis, University of Heidelberg (2009). 

\bibitem{markThesis} M.~Heisel, \emph{LArGe -- A liquid argon scintillation
 veto for GERDA}, Ph.~D. thesis, University of Heidelberg (2011). 

\bibitem{SQLref} SQL - Structured Query Language - Standard ISO/IEC 9075.
\end{thebibliography}
\end{document}